\documentstyle[seceq,preprint,epsf]{ptptex}
\newcommand{\beq}{\begin{equation}}
\newcommand{\beqa}{\begin{eqnarray}}
\newcommand{\eeq}{\end{equation}}
\newcommand{\eeqa}{\end{eqnarray}}
\newcommand{\bm}[1]{\mbox{{\boldmath $#1$}}} 
\def\siml{\mathrel{\mathpalette\Oversim<}}
\def\simg{\mathrel{\mathpalette\Oversim>}}
\def\Oversim#1#2{\lower0.5ex\vbox{\baselineskip0pt\lineskip0pt%
            \lineskiplimit0pt\ialign{%
          $\mathsurround0pt #1\hfil##\hfil$\crcr#2\crcr\sim\crcr}}}
\newcommand{\bfeta}{\mbox{\boldmath $\eta$}}
\newcommand{\bfxi}{\mbox{\boldmath $\xi$}}
\newenvironment{namelist}[1]{%
  \begin{list}{}
    {
      \settowidth{\labelwidth}{#1}
      \setlength{\leftmargin}{1.1\labelwidth}}
    }{%
    \end{list}}

\title{%        %You can use \\ for explicit line-break
Kepler Rotation Effects\\ on the Binary-Lens Microlensing Events
}
\author{%       %Use \sc for the family name
  Kunihito {\sc Ioka},$^{1}$
  Ryoichi {\sc Nishi},$^{1}$
  and Yukitoshi {\sc Kan-ya}$^{2}$
}

\inst{%         %Affiliation, neglected when [addenda] or [errata]
$^{1}$Department of Physics, Kyoto University, Kyoto 606-8502, Japan
\\
$^{2}$National Astronomical Observatory, Mitaka, Tokyo, 181-0015, Japan
}

\recdate{%      %Editorial Office will fill in this.
May, 24, 1999
}

\abst{%       %this abstract is neglected when [addenda] or [errata]
  We investigate the effects of the Kepler rotation of lens binaries
  on the binary-microlensing events towards the Large Magellanic
  Cloud (LMC) and the Small Magellanic Cloud (SMC).
  It is found that the rotation effects cannot always be neglected
  when the lens binaries are in the LMC disk or the SMC disk, i.e.,
  when they are self-lensing.
  Therefore we suggest that it will be necessary to consider
  the rotation effects
  in the analyses of the coming binary events
  if the microlensing events towards the halo are self-lensing.
  As an example, we reexamine the MACHO LMC-9 event,
  in which the slow transverse velocity of the lens binary
  suggests a microlensing event in the LMC disk.
  From a simple analysis, it is shown that the lens binary
  with total mass $\sim 1M_{\odot}$ rotates by more than $\sim 60^{\circ}$
  during the Einstein radius crossing time.
  However, the fitting of MACHO LMC-9 with an additional parameter,
  the rotation period, shows that the rotation effects are small,
  i.e., the projected rotation angle is only 
  $\sim 5.9^{\circ} (M/M_{\odot})^{1/4}$ 
  during the Einstein radius crossing time.
  This contradiction can be settled if the physical parameters,
  such as the mass and the velocity, are different in this event,
  the binary is nearly edge-on, or the binary is very eccentric,
  though definite conclusions cannot be drawn from this single event.
  If the microlensing events towards the halo are due to self-lensing,
  binary-events for which the rotation effects are important will increase
  and stronger constraints on the nature of the lenses will be obtained.
}

\begin{document}

\maketitle

\section{Introduction}
The analysis of the first $2.1$ years of photometry of $8.5 \times 10^6$
stars in the LMC by the MACHO Collaboration
\cite{alcock} suggests that the fraction $0.62^{+0.3}_{-0.2}$
of our halo consists of massive compact halo objects (MACHOs)
of mass $0.5^{+0.3}_{-0.2}
M_{\odot}$ in the standard spherical flat rotation halo model.
A preliminary analysis of four years of data suggests the existence of 
at least eight additional microlensing events with $t_{\rm dur}\sim 90$ days
in the direction of the LMC. \cite{cook}

At present, we do not know what MACHOs are.
%There are several candidates proposed to explain MACHOs,
There have been several identifications of MACHOs proposed,
such as brown dwarfs, red dwarfs, white dwarfs, neutron stars,
primordial black holes, and so on.
\cite{cook,bahcall,flynn,GFB98,graff,ryu,chabrier,chabrier2,adams,fields,charlot,gibson,canal,honma,yokoyama,kawasaki,jedamzik,bhmacho,bhmacho2,bhmacho3}
Any objects clustered somewhere between the LMC and the Sun 
with column density larger than  $25M_{\odot}{\rm pc}^{-2}$
may also explain the data. \cite{naka96}
They include the following possibilities:
LMC-LMC self-lensing, the spheroid component, thick disk, 
a dwarf galaxy, tidal debris, and warping and flaring of 
the galactic disk. \cite{sahu,wu,zhao97,evans97,gate97,BE98,AU99}
(See also Ref. \citen{NI99}.)

Such obscurities of the mass and the spatial distribution essentially
result from the fact that the time scale of an event,
which is an important observable,
is a degenerate combination of the three quantities
one would like to know,
the mass, the velocity and the position of the lensing object.
Several methods have been proposed to break these degeneracies,
for example, launching a parallax satellite into solar orbit,
\cite{GO94sa,GO95,BO96,GA97,HO99sa}
observing the annual modulation in light magnification induced by the
Earth's motion, \cite{GO92,GO98para}
observing the deviation of light magnification
from a simple point-source model due to the
finite-source size effect
when the impact parameter of the trajectory of the
lens is comparable to the source size, \cite{GO94pr,NE94,WI94}
distinguishing the dependence of the lensing rate on the background
stellar density, \cite{st98}
and so on.

A microlensing event due to a binary is one of the best candidates
to break the degeneracies.
In binary-microlensing events,
the light magnification dramatically deviates from 
that of a simple point-source model
when the source transverses the caustics,
where a point source is amplified infinitely.
We can obtain information concerning the transverse velocity of the source
from this deviation,
which can be used to distinguish 
between halo-lensing and self-lensing.
To this time, two binary-lens microlensing events have been observed,
MACHO LMC-9 \cite{bennett} and MACHO 98-SMC-1.
\cite{AF98,ALB98,ALC98}
Although we cannot say for certain from only these events, \cite{HO99}
these events support self-lensing because of slow transverse velocities.
In the future, the number of binary-lens microlensing events will increase,
\cite{MA91,BO94}
and hence these events will be important to break 
the degeneracies in the physical parameters.

In almost all analyses of binary-lens microlensing events,
the rotation of the lens binary has been neglected.\cite{bennett}
This is because the period of a lens binary in the halo
is much larger than the time scale of the amplification.
However, the lensing object of the MACHO LMC-9 event, for example,
is very likely to reside in the LMC disk, not in our halo.
Since the characteristic transverse velocity of a lens
in the LMC disk or the SMC disk is smaller than that in our halo,
the rotation of the lens binary in the disk may be important.
For this reason we reconsider the rotation effects
on the analyses of the binary-lens microlensing events in this paper.
As an example we reanalyze the MACHO LMC-9 event
taking the rotation into account.
Note that, in the previous analysis of this event,
the transverse velocity is somewhat
smaller than that expected for a lens in the LMC disk.\cite{bennett}
If we take the rotation of the lens binary into account,
the transverse velocity may be larger,
since the incident angle of the trajectory of the source into the
caustics may be smaller, and hence the source may take shorter time
to move by one stellar radius of the source.
This is one of our motivations to examine the rotation effects.

In \S 2 microlensing by a double point mass is reviewed.
In \S 3 the rotation effects of a lens binary are estimated.
We suggest the possibility that the rotation effects are important 
when the lens binary resides in the LMC disk or the SMC disk.
In \S 4 we perform the fitting of the MACHO LMC-9 event, taking into
account the rotation of the lens binary.
Section 5 is devoted to summary and discussion.

\section{Microlensing by two point masses}
\begin{figure}[h]
  \epsfysize 8cm \centerline{\epsfbox{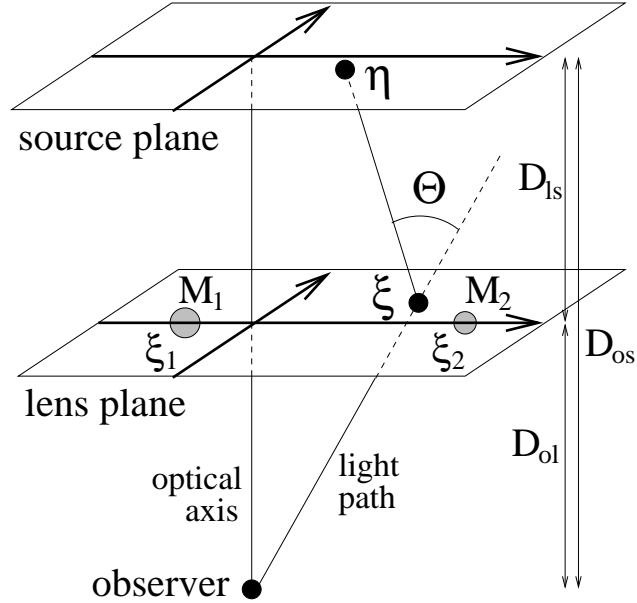}}
  \caption{The geometry of the gravitational lensing by a double point 
    mass lens.}
  \label{fig:geom}
\end{figure}

We now briefly review microlensing by a double point mass
\cite{schneider,SEF} to introduce our notation.
We consider a lens binary consisting of two point masses, $M_1$ and $M_2$,
whose center is
at a distance $D_{ol}$ from the observer,
and we consider a source at a distance $D_{os}$ from the observer
in Fig.~\ref{fig:geom}.
We define the lens plane as the plane which contains the center of mass
of the lens binary and
is perpendicular to the line connecting the observer and the center of
mass, i.e., the optical axis.
We also define the source plane as the plane which contains the source and is
parallel to the lens plane.
The distance between the lens plane and the source plane is written as
$D_{ls}=D_{os}-D_{ol}$.
We define a coordinate system $(\xi_x,\xi_y)$ on the lens plane and
$(\eta_x,\eta_y)$ on the source plane,
taking the origin of each coordinate at the intersection between each
plane and the optical axis.
An equation which relates the image position $\bfxi$
to the source position $\bfeta$ is
called a ``lens equation''.
From Fig.~\ref{fig:geom}, 
we see that the lens equation for lensing by a double point mass is
\beqa
\bfeta&=&{{D_{os}}\over{D_{ol}}}{\bfxi} -{D_{ls}}{\bm \Theta} ({\bfxi}),
\label{lenseq}
\\
{\bm \Theta}({\bfxi})&=&{{4 G M_1}\over{c^2}}
{{{\bfxi}-{\bfxi}_1}\over{|{\bfxi}-{\bfxi}_1|^2}}
+{{4 G M_2}\over{c^2}}
{{{\bfxi}-{\bfxi}_2}\over{|{\bfxi}-{\bfxi}_2|^2}},
\label{dangle}
\eeqa
where ${\bfxi}_1$ and ${\bfxi}_2$ are the positions of the masses
projected onto the lens plane.
${\bm\Theta}({\bfxi})$ is the deflection angle of light due
to the lens masses,
which is the summation of the deflection angle due to each mass.
The Einstein radius for the total mass of the binary, $M=M_1+M_2$, is
defined as
\beq
\rho_E:=\sqrt{{{4GM}\over{c^2}}{{D_{ol} D_{ls}}\over{D_{os}}}}
=\sqrt{{{4GMD_{os}}\over{c^2}} x(1-x)},
\label{defEradi}
\eeq
where $x:={{D_{ol}}/{D_{os}}}$.
With the definitions
\beq
{\bm r}:={{\bfxi}\over{\rho_E}},\quad
{\bm z}:=x {{\bfeta}\over{\rho_E}},\quad
\mu_i:={{M_i}\over M}, \quad{(i=1,2)}
\label{dlessdef}
\eeq
the lens equations (\ref{lenseq}) and (\ref{dangle}) become
dimensionless equations,
\beqa
{\bm z}&=&{\bm r}-{\bm \Theta}({\bm r}),
\label{leq}
\\
{\bm \Theta}({\bm r})&=&
\mu_1 {{{\bm r}-{\bm r}_1}\over{|{\bm r}-{\bm r}_1|^2}}
+\mu_2 {{{\bm r}-{\bm r}_2}\over{|{\bm r}-{\bm r}_2|^2}}.
\label{leq2}
\eeqa
Note that
we are discussing the situation on the lens plane,
since we normalize the length as in Eq.~(\ref{dlessdef}).

\begin{figure}[t]
\begin{center}  
  \begin{tabular}[h]{ccc}
    \epsfysize 5cm 
    \epsfbox{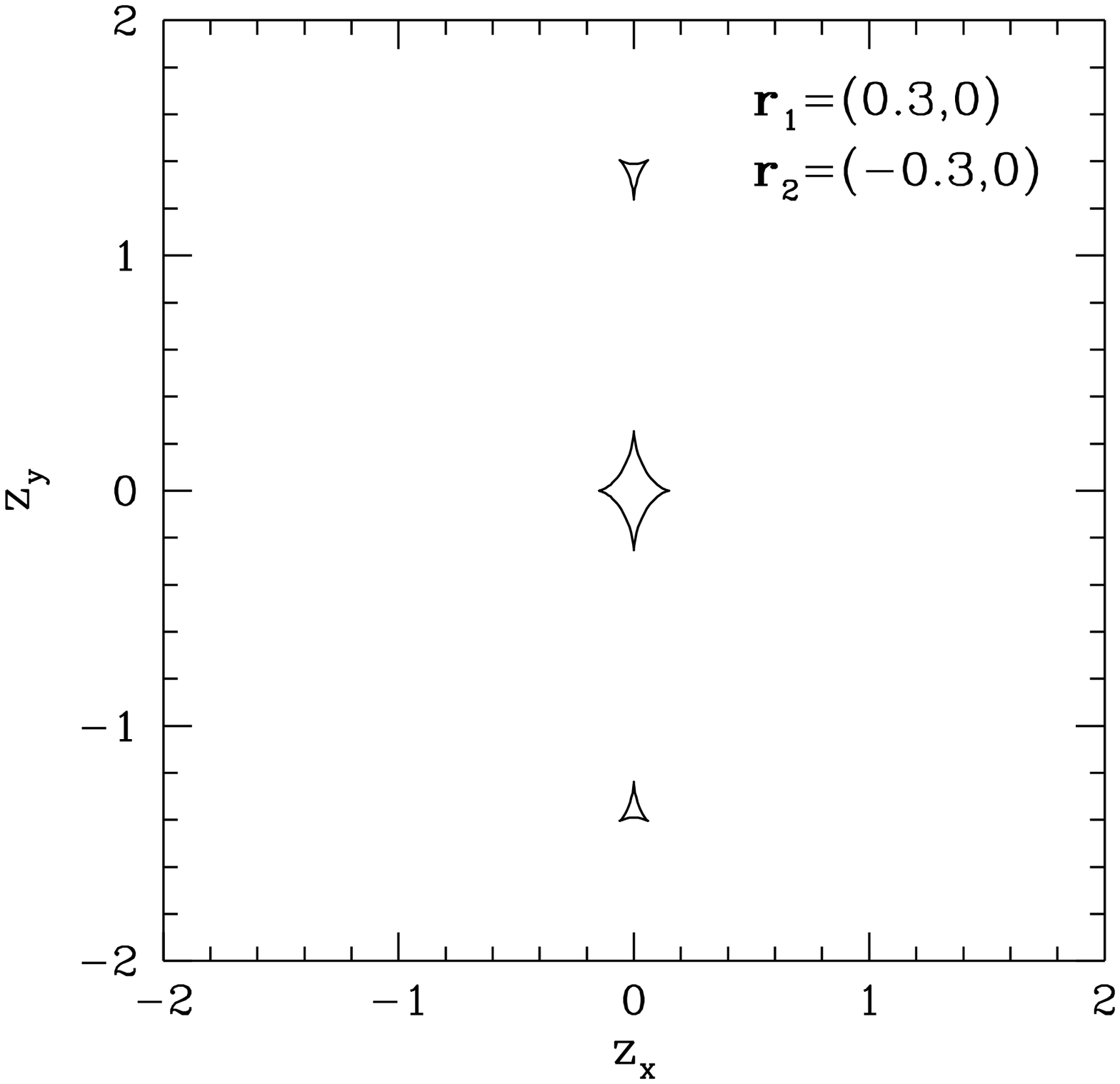} &
    \epsfysize 5cm 
    \epsfbox{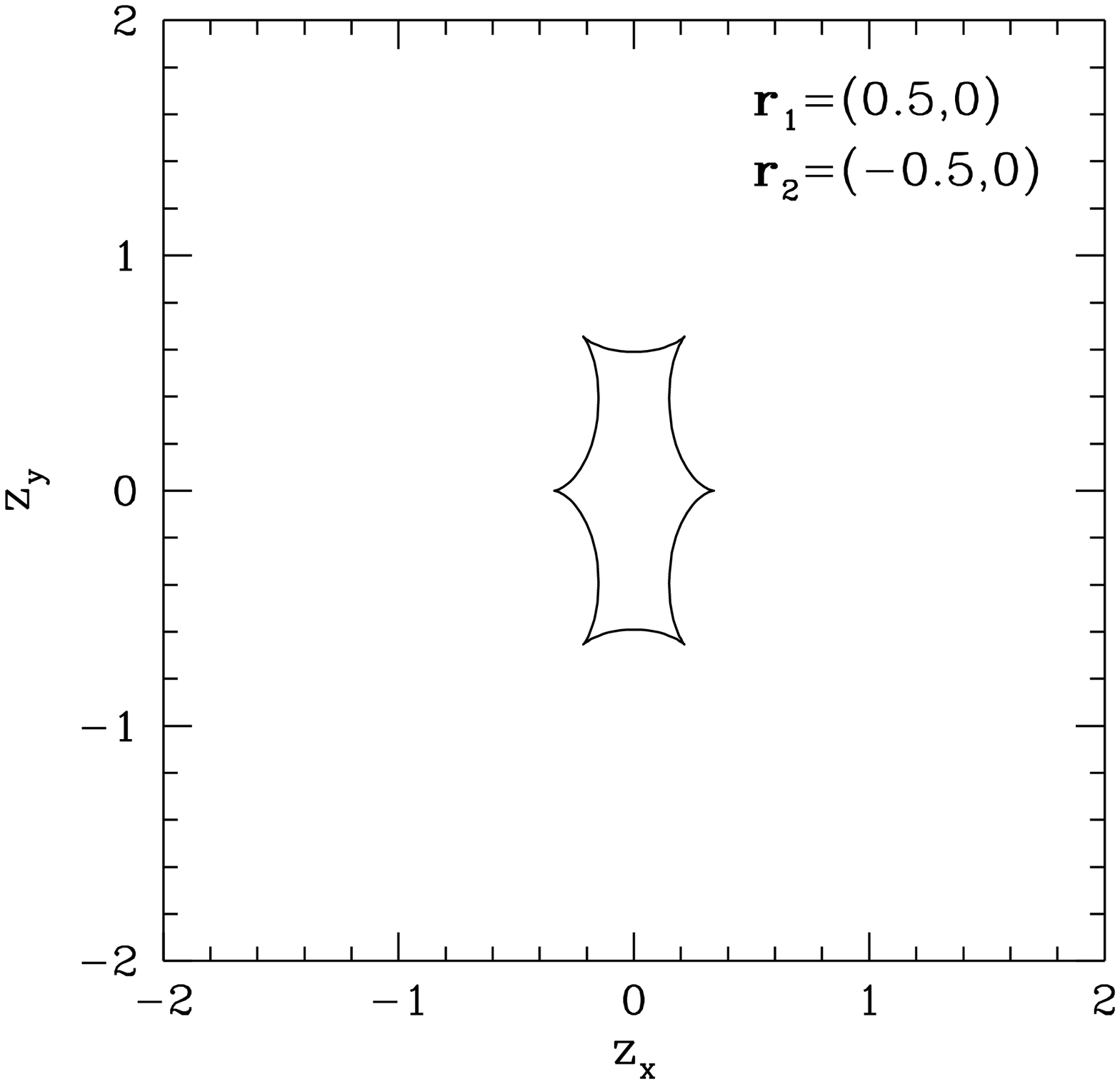} &
    \epsfysize 5cm 
    \epsfbox{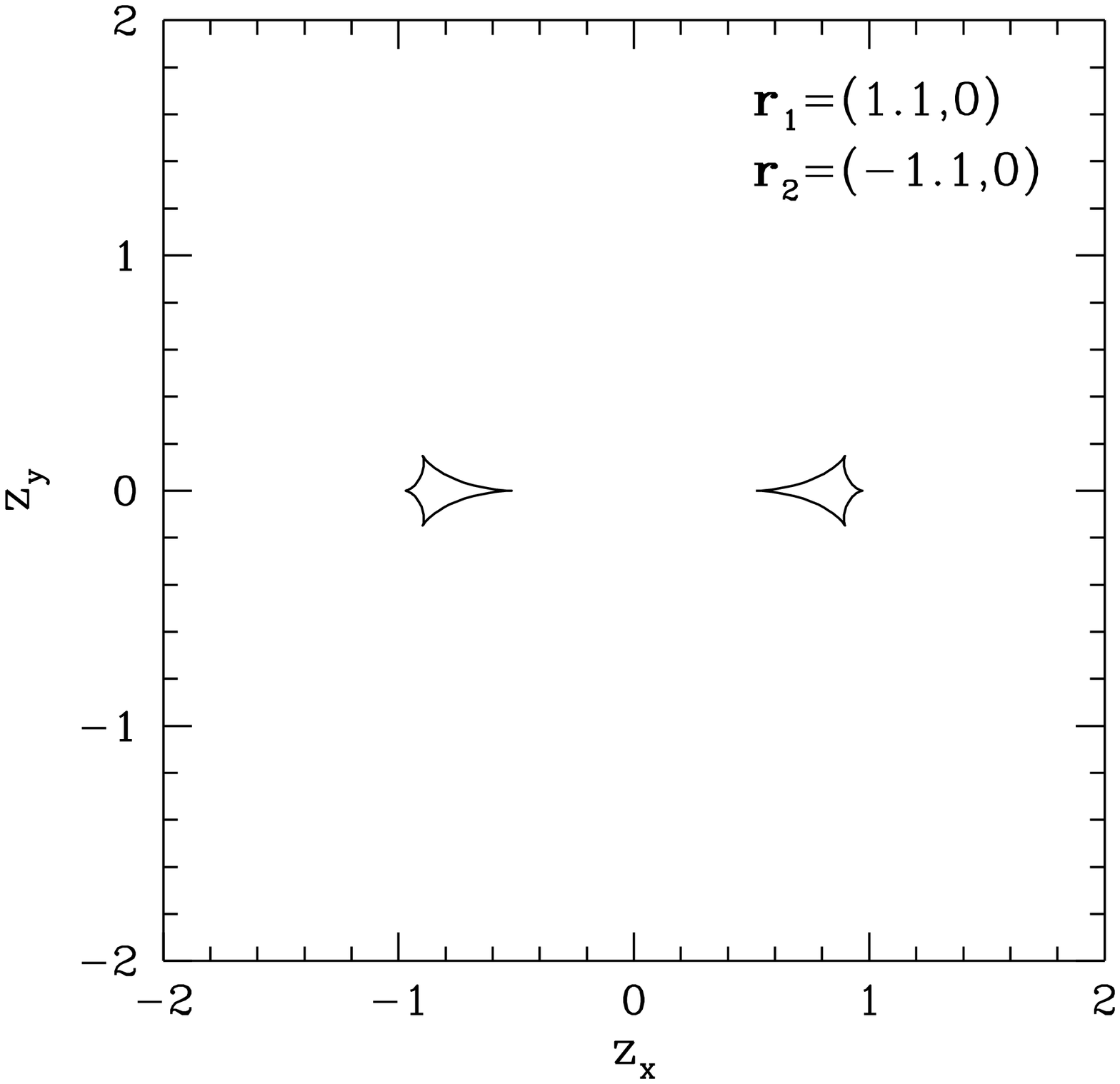}
  \end{tabular}
\end{center}    
  \caption{Caustics for a double point lens.
    The lenses have identical masses 
    and are separated by $l=0.6,1.4$ and $2.2$.}
  \label{fig:caus}
\end{figure}

These equations (\ref{leq}) and (\ref{leq2}) 
can be used to find all images of the source. \cite{witt}
If we cannot resolve the images by observations,
the only observable quantity is the amplification of the brightness
of the source.
The amplification factor $I$ is the inverse of the determinant of the
Jacobi matrix,
\beq
I=\sum_i \left.\left|\det\left({{\partial {\bm z}}\over{\partial {\bm
          r}}}\right)\right|^{-1}\right|_{{\bm r}={\bm r}_i},
\eeq
where ${\bm r}_i$ is the image position.
For certain values of ${\bm r}$, the amplification factor $I$
diverges.
A set of these points forms curves called ``critical curves''.
The projection of the critical curves onto the source plane with
Eqs.~(\ref{leq}) and (\ref{leq2}) forms caustics on the source plane.
Caustics have three kinds of morphology, depending on the separation of
the masses.
In Fig.~\ref{fig:caus} these three kinds of caustics are shown.
The number of images is five in the closed caustics and three outside.
Around the caustics, large amplification appears, and the light curve
has a peak.

\section{Microlensing events by a binary lens}
\subsection{Motion of the source}
We assume that we can neglect the Earth's motion around the Sun, and
consider the relative motion of the observer, the
lens and the source as the motion of the source.
Every quantity on the source plane is projected onto the lens plane
by Eq.~(\ref{dlessdef}).
Hence it is convenient to consider the motion of the source as that
projected onto the lens plane with Eq.~(\ref{dlessdef}).
The trajectory of the source is characterized by its
impact parameter $b$ with respect to the origin of the lens plane
in units of the Einstein radius $\rho_E$ and the angle $\theta$
between the $x$-axis on the lens plane and the trajectory.
The source closest approaches the origin at $T_S$,
and the transverse velocity of the source in the lens plane is $V_\bot$.
The transverse velocity of the source pulled back onto the source plane
$V_T$ is related to $V_\bot$ as
\beq
V_T={{V_{\bot}}\over{x}}.
\label{defVT}
\eeq
The Einstein radius crossing time $t_E$ is defined as
\beq
t_E:={{\rho_E}\over{V_{\bot}}}.
\label{deftE}
\eeq
It is convenient to measure the time $T$ with respect to $t_E$,
$t:={{T}/{t_E}}$.
By normalizing the time in this manner,
the source moves by a unit length in a unit time.
The source closest approaches the origin at the normalized time,
$t_S: =T_S/t_E$.

If there are enough data around the peak of the light curve,
important information about the size of the source can be obtained.
\cite{bennett}
If the radius of the source is $R_\star$, it takes a time
\beq
T_\star={{R_\star}\over{V_T}}=x {{R_\star}\over{V_\bot}}
\label{defTstar}
\eeq
for the source to move by one stellar radius of the source.

\subsection{Rotation of the lens binary}
The lens binary rotates according to Kepler's law.
The relative vector ${\bm U}$
is defined by ${\bm U}:={\bm U}_1-{\bm U}_2$,
where ${\bm U}_1$ and ${\bm U}_2$ denote the positions of the binary
masses in the orbital plane.
We take the origin of the orbital plane at the 
center of the binary masses as
$M_1 {\bm U}_1 + M_2 {\bm U}_2 =0$.
Hence, the positions of the binary masses are determined by the relative
vector ${\bm U}$ as ${\bm U}_1={\mu_2} {\bm U}$ and
${\bm U}_2=-{\mu_1} {\bm U}$,
where $\mu_1$ and $\mu_2$ are defined in Eq.~(\ref{dlessdef}).
In the orbital plane, the relative vector
${\bm U}=(U\cos \phi, U\sin \phi)$ at a time $T$ is determined through the
parameter $\lambda$ as
\beqa
T-T_0&=&{{T_B}\over{2\pi}}(\lambda - e\sin\lambda),
\label{bTime}
\\
U&=&A (1-e\cos \lambda),
\label{bLength}
\\
\cos (\phi -\phi_0)&=& {{\cos \lambda -e }\over{1 - e \cos \lambda}},
\label{bAngle}
\eeqa
where
\beq
T_B=2\pi\sqrt{{{A^3}\over{GM}}}
\label{period}
\eeq
is the period,
$A$ is the
semi-major axis, and $e$ is the eccentricity of the lens binary.
$T_0$ and $\phi_0$ are integral constants.
We can take the coordinates on the orbital plane so that $\phi_0 =0$.
With the definitions
\beqa
t_B:={{T_B}\over{t_E}},\quad
t_0:={{T_0}\over{t_E}},\quad
{\bm u}:={{\bm U}\over{\rho_E}},\quad
a:={{A}\over{\rho_E}},
\label{deftB}
\eeqa
the above equations (\ref{bTime}), (\ref{bLength}) and
(\ref{bAngle}) become dimensionless:
\beqa
t - t_0 &=& {{t_B}\over{2\pi}}(\lambda - e\sin\lambda),
\label{btime}
\\
u&=&a (1-e\cos \lambda),
\label{blength}
\\
\cos \phi&=& {{\cos \lambda -e }\over{1 - e \cos \lambda}}.
\label{bangle}
\eeqa

\begin{figure}[t]
  \epsfysize 8cm \centerline{\epsfbox{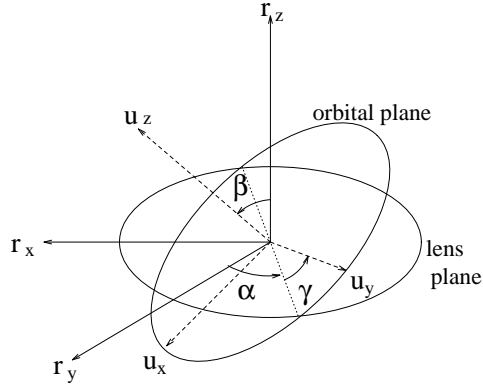}}
  \caption{The relation between the orbital plane and the lens
plane.
The $(r_x, r_y)$ plane is the lens plane,
the $(u_x, u_y)$ plane is the orbital
plane, and $\alpha$, $\beta$ and $\gamma$ are Euler angles.}
  \label{fig:euler}
\end{figure}

In general, the orbital plane does not coincide with the lens
plane.
Since we consider the case $D_{ol} \gg A$,
it is convenient to think about the lens
masses projected onto the lens plane, i.e., the thin lens approximation.
The orientation of the coordinates in the orbital plane
relative to
the coordinates in the lens plane is determined by Euler angles
$\alpha$, $\beta$ and $\gamma$.
In Fig.~\ref{fig:euler}, 
we show the relation between the orbital plane and the lens
plane, where
$(r_x, r_y)$ plane is the lens plane,
$(u_x, u_y)$ plane is the orbital
plane, and $\alpha$, $\beta$ and $\gamma$ are Euler angles.
We can take the coordinates in the lens plane so that $\alpha = 0$,
since $\alpha$ can be absorbed into $\theta$, the angle
between the $x$-axis in the lens plane and the trajectory
of the source.
A point $(u \cos \phi, u \sin \phi)$
in the orbital plane is projected
to the point $({l} \cos \varphi, {l} \sin \varphi)$
in the lens plane, where
\beqa
l &=& u \sqrt{\cos^2\beta \cos^2(\phi+\gamma) +
\sin^2(\phi+\gamma)},\label{proj1}\\
\tan \varphi &=& {{\tan (\phi + \gamma)}\over{\cos \beta}}.
\label{proj2}
\eeqa

%The rotation of the binary appears as
%two effects.
%One effect is the deformation of the trajectory of the source,
%since we can consider the relative motion of the lens and the source
%as the motion of the source.
%The other is the change of the separation $l$
%between the projected lens masses.

\subsection{Estimate of the rotation effects}\label{sec:rote}
We can estimate the rotation effects of a lens binary
by comparing the period of the binary $T_B$
with the Einstein radius crossing time $t_E$.
If the ratio $t_B={{T_B}/{t_E}}$, is not much larger than unity,
we cannot neglect the rotation of the binary.

For a typical MACHO, the ratio $t_B$ is given by
\begin{eqnarray}
  &&t_B = {{T_B}\over{t_E}} = {{2\pi V_{\perp}}\over{\rho_E}}
  \sqrt{{A^3}\over{GM}}
  = {{2\pi V_{\perp}}\over{\rho_E}}
  \sqrt{{\rho_E^3 a^3}\over{GM}}
  \nonumber\\
  &\sim& 134 \left({{a}\over{1}}\right)^{3/2}
  \left( {{D_{OS}}\over{50\ {\rm kpc}}} \right)^{1/4}
  \left({{M}\over{M_{\odot}}}\right)^{-1/4}
  \left({{V_{\perp}}\over{200\ {\rm km/s}}}\right)
  \left({{x(1-x)}\over{0.5(1-0.5)}}\right)^{1/4},
  \label{halotB}
\end{eqnarray}
using Eqs.~(\ref{defEradi}), (\ref{deftE}), (\ref{period})
and (\ref{deftB}).
This implies that a typical binary in our halo rotates by
about $360^{\circ}/t_B \simeq 2.7^{\circ} (M/M_{\odot})^{1/4}$
during the
Einstein radius crossing time.
This may be small enough to neglect the rotation of the binary.\footnote{
For a close binary, i.e., for $a\ll 1$, it seems that the 
rotation effects can be large from Eq.~(\ref{halotB}).
However, this is not a correct argument, as discussed
in Appendix \ref{sec:close}.}

On the other hand, if a lens binary resides in the LMC disk,
the parameters are different.
For a typical LMC lens, assuming that the thickness of the LMC
disk is smaller than $\sim 500$ pc, the ratio $t_B$ is given by
\begin{equation}
  t_B \siml 18 \left({{a}\over{1}}\right)^{3/2}
  \left( {{D_{OS}}\over{50\ {\rm kpc}}} \right)^{1/4}
  \left({{M}\over{M_{\odot}}}\right)^{-1/4}
  \left({{V_{\perp}}\over{60\ {\rm km/s}}}\right)
  \left({{x(1-x)}\over{0.99(1-0.99)}}\right)^{1/4},
  \label{tBLMC1}
\end{equation}
which implies that the binary rotates by more than
$\sim 20^{\circ} (M/M_{\odot})^{1/4}$ during the Einstein radius crossing time.
Thus, if the lens resides in the LMC disk, the rotation effects 
may be important for the fitting of the light curve.
This argument can also be applied to an SMC lens, since the transverse
velocity is small and $x$ is close to 1.

Bennett et al. \cite{bennett} fitted
the data of MACHO LMC-9 and obtained
fitted parameters as in Table \ref{tab:MACHO9byMACHO}.
The radius of the source is estimated as
\begin{equation}
  R_\star=1.5 \pm 0.2 \ {\rm R_{\odot}},
  \label{defRstar}
\end{equation}
using the theory of stellar evolution.
The transverse velocity in the source plane can be estimated as
\begin{equation}
  V_T = {{R_\star}\over{T_\star}} = 19 \pm 6 \ {\rm km/s},
  \label{VT9}
\end{equation}
with $T_\star$ in Table \ref{tab:MACHO9byMACHO} and
Eq.~(\ref{defTstar}).
Comparing this value with the probability distribution of the transverse 
velocity in the LMC disk and the Milky Way halo,
a lens in the LMC disk is preferred over a halo lens.\cite{bennett}
Assuming that the lens of MACHO LMC-9 resides in the LMC disk,
the ratio $t_B$ is given by
\begin{equation}
  t_B < 5.7 \left({{a}\over{1}}\right)^{3/2}
  \left( {{D_{OS}}\over{50\ {\rm kpc}}} \right)^{1/4}
  \left({{M}\over{M_{\odot}}}\right)^{-1/4}
  \left({{V_{\perp}}\over{19\ {\rm km/s}}}\right)
  \left({{x(1-x)}\over{0.99(1-0.99)}}\right)^{1/4},
  \label{tB9}
\end{equation}
where we use the relation $V_\bot < V_T$, since $V_\bot = x V_T$ and $x<1$.
Since Eq.~(\ref{tB9}) indicates that the binary rotates by more than $\sim
60^{\circ} (M/M_{\odot})^{1/4}$ during the Einstein radius crossing time,
we cannot neglect the rotation of the binary for the fitting of the data.
Therefore we perform the fitting of the MACHO LMC-9 event 
taking into account the rotation of the lens binary in the next section.

The conclusion of this section is that
the rotation effects cannot be always neglected
when the lens binaries are in the LMC disk or the SMC disk,
while the rotation can be neglected 
when the lenses are in the Milky Way halo.

\section{Fitting of the observed data in the binary-lens events}
\subsection{Parameters characterizing the binary-lens
  microlensing events}
\label{sec:para}
We summarize the parameters necessary
to describe the binary-lens microlensing event in this section.
First, a dual color observation requires two parameters,
taking the blending into consideration:
\begin{itemize}
\item[(1)] $f_{oR}$ : fraction of the lensed brightness of the lensed star
  in the red band.
\item[(2)] $f_{oB}$ : fraction of the lensed brightness of the lensed star
  in the blue band.
\end{itemize}
When we can neglect the rotation of the lens binary,
we need the following parameters:
\begin{itemize}
\item[(3)] $t_E$ : Einstein radius crossing time.
\item[(4)] $T_S$ : time when the source most closely approaches the origin.
\item[(5)] $b$ : impact parameter of the source in the
  lens plane in units of the Einstein radius.
\item[(6)] $\theta$ : angle between the $x$-axis in the lens
  plane and the trajectory of the source.
\item[(7)] $\mu_1$ : the mass fraction of the first mass.
\item[(8)] $l$ : separation of the lens masses
  projected onto the lens plane in units of the Einstein radius.
\item[(9)] $T_\star$ : time for the source to move by one stellar radius of
  the source.
\end{itemize}
If we can neglect the finite size of the source,
the last parameter
$T_\star$ is not necessary.

When we consider the rotation of the lens binary,
we need the following
parameters in addition to the above parameters:
\begin{itemize}
\item[(8)] $a$ : semi-major axis of the lens binary in units of the 
  Einstein radius.
\item[(10)] $t_B$ : period of the lens binary in units of the
  Einstein radius crossing time.
\item[(11)] $\beta$ : Euler angle.
\item[(12)] $\gamma$ : Euler angle.
\item[(13)] $e$ : eccentricity of the lens binary.
\item[(14)] $T_0$ : time when the binary is at the pericenter.
\end{itemize}
where the parameter (8) has been replaced.
We need five additional
parameters when we take the rotation of the lens binary into account.

If we assume that the orbit of the binary is circular ($e=0$),
the parameters (13) and (14) are not necessary,
since $T_0$ can be absorbed into $\gamma$.
If we assume that the orbit of the binary is face-on,
the parameters (11) and (12) are not necessary,
since $\gamma$ can be absorbed into $\theta$.
If we assume that the lens binary is face-on and $e=0$,
the parameters from (11) to (14) are not necessary.

\subsection{Fitting of the MACHO LMC-9 event}\label{sec:fitting}
We analyze the raw data for the MACHO LMC-9 event.
The baseline of the photometry that corresponds to no amplification
has to be determined by the fitting with it added as
one more parameter.
However, to save time, we determine the baseline by
the least squares fitting of the data
for $T<300$ day and $900$ {\rm day} $< T$, where
the amplification is expected to be less than 
$(4^2+2)/(4\times \sqrt{4^2+4}) \sim 1.006$.
With this baseline we can translate the raw data into
the data for the amplification.

\begin{figure}[t]
\begin{center}  
  \begin{tabular}[h]{ccc}
    \epsfysize 5cm 
    \epsfbox{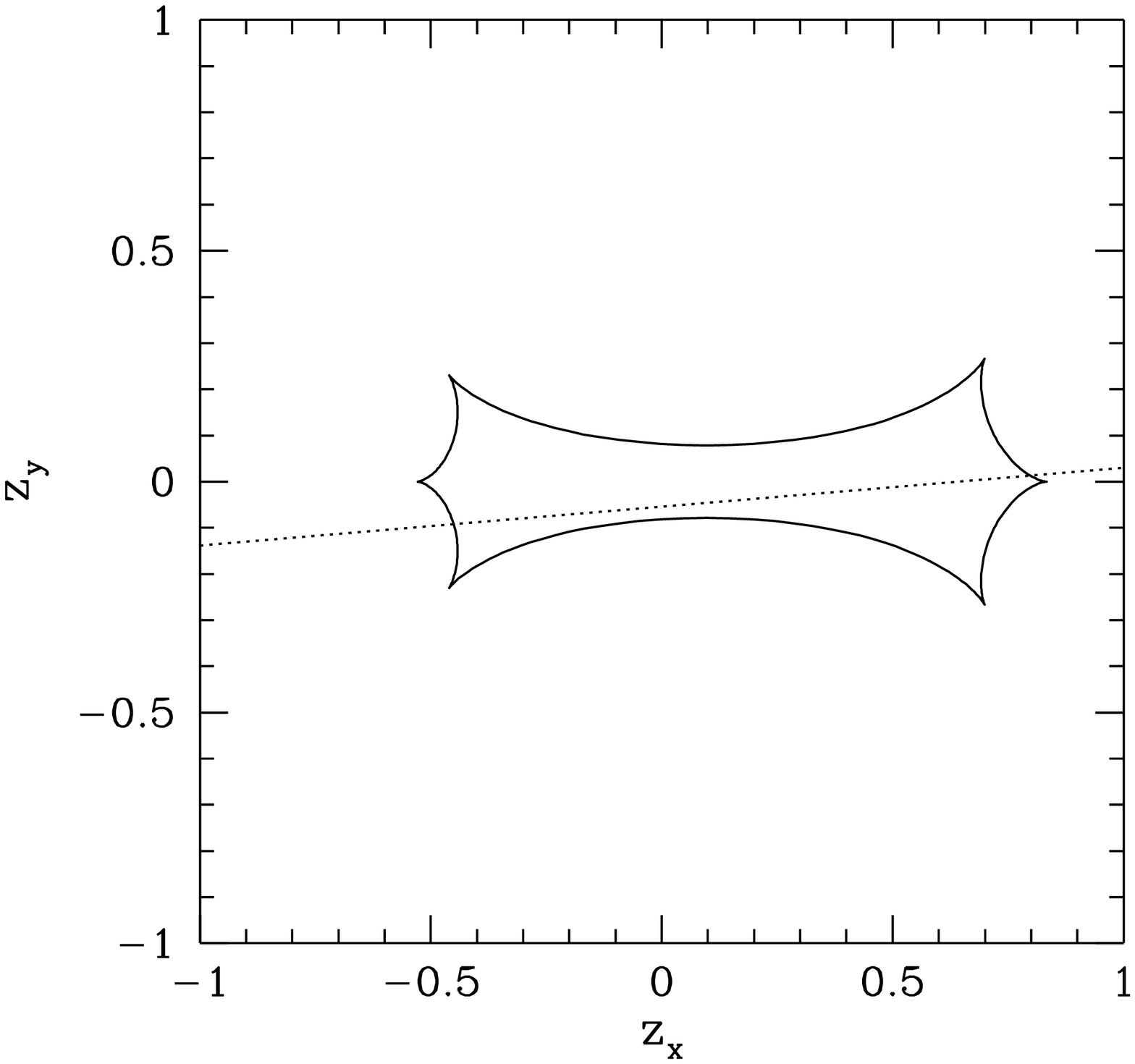} &
    \epsfysize 5cm 
    \epsfbox{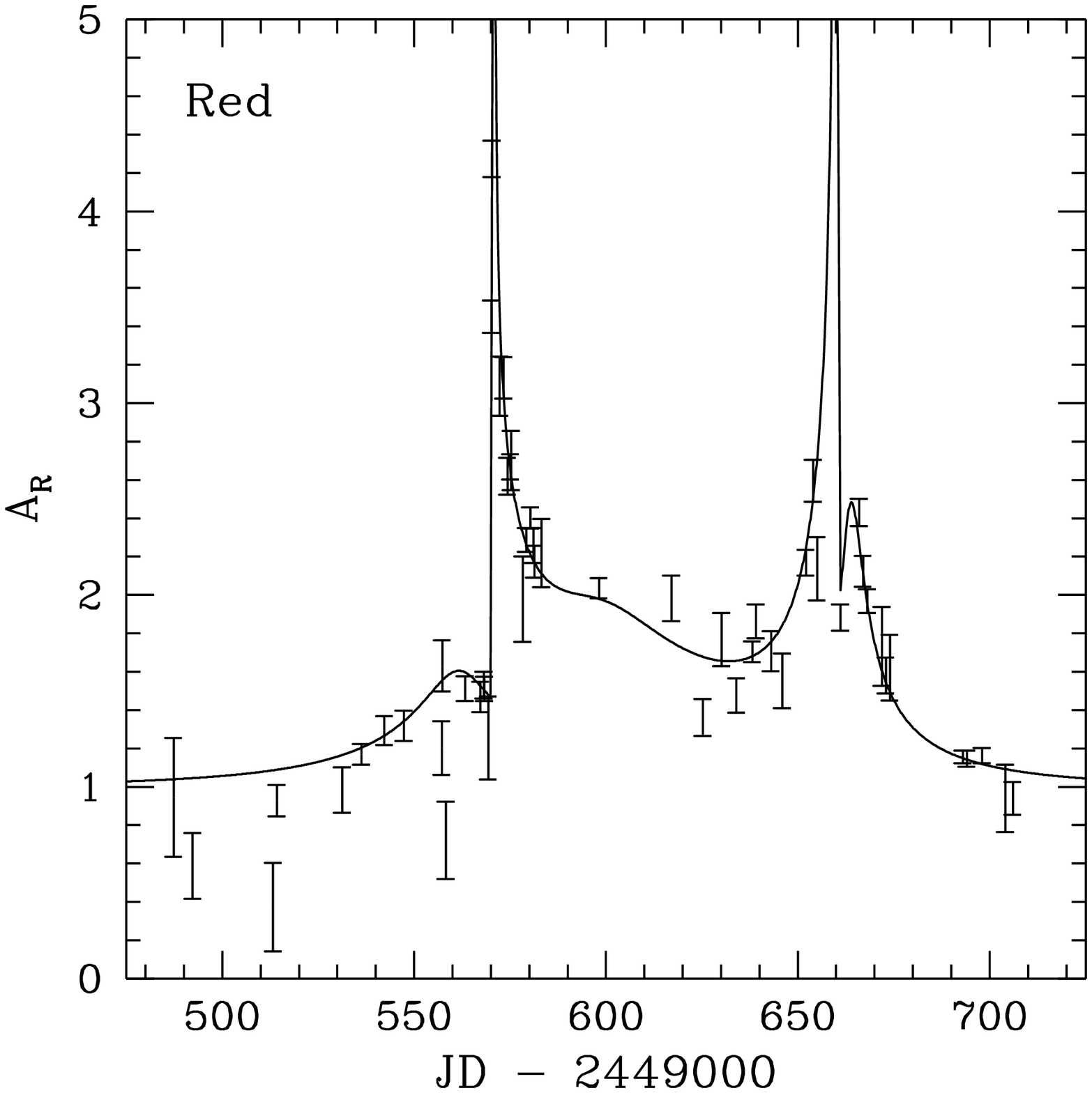} &
    \epsfysize 5cm 
    \epsfbox{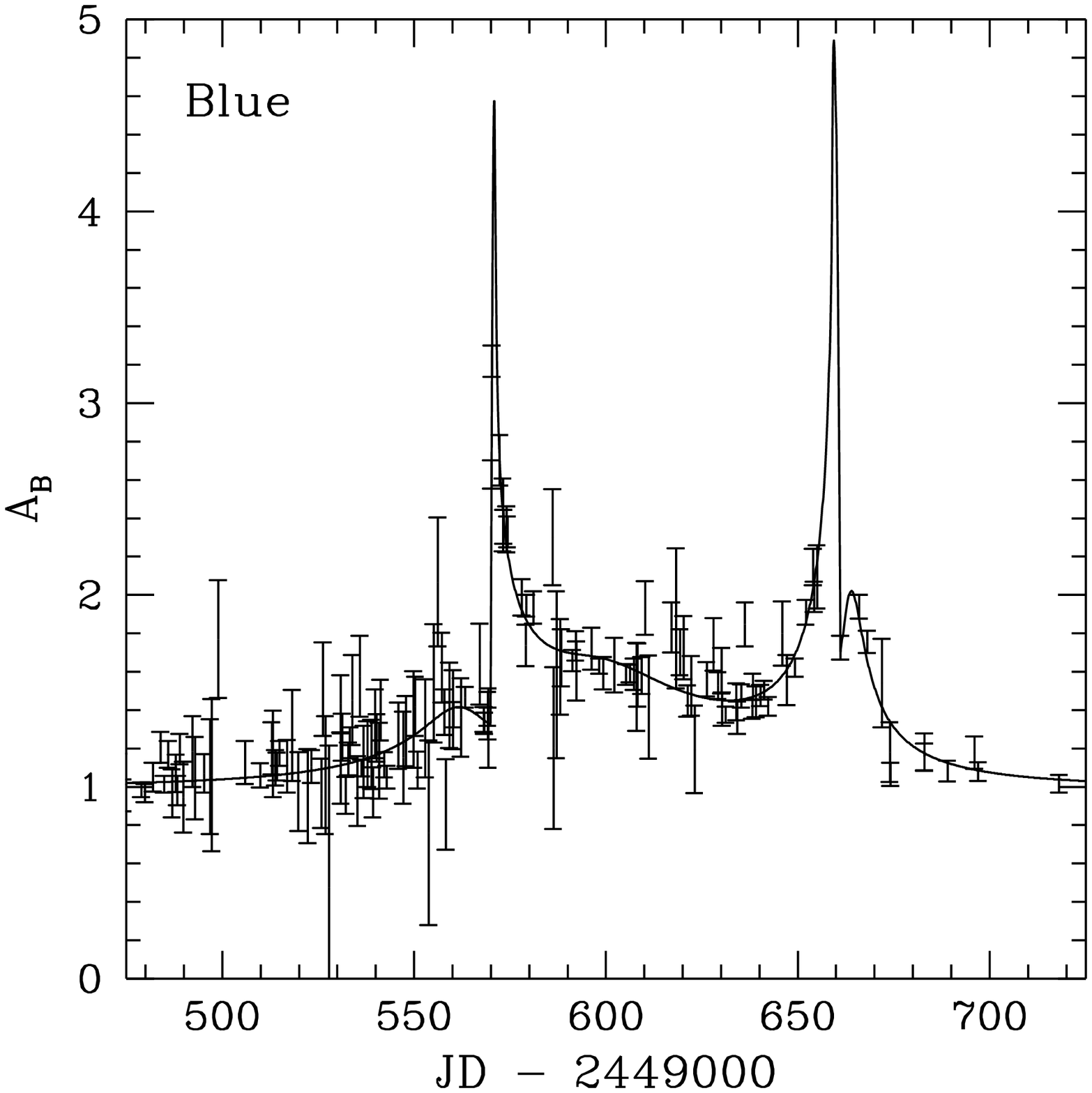}
  \end{tabular}
\end{center}    
  \caption{The structure of the caustics and the trajectory of the source
are shown ({\it left}).
The corresponding light curves for red ({\it middle})
and blue ({\it right}) are also shown.
The rotation effects are not included.}
  \label{fig:static}
\end{figure}

We first performed the fitting of the data of MACHO LMC-9 neglecting the
rotation of the lens binary to check our fitting code.
The result of the $\chi^2$ fit is shown in Table
\ref{tab:nonrotatingMACHO9}.\footnote{
There are generally
several local minima in the $\chi^2$ fitting using some
parameters.
Several methods have been proposed to find the global minimum.
\cite{MaoStefano,DSP97,ALB99}
However, this requires a great deal of effort to find the global minimum.
Therefore, we restrict the parameter space somewhat from 
the shape of the light curve, and then
we pick the parameter set with the smallest $\chi^2$
from $\sim 50$ fittings.
For example, we can restrict the morphology of the caustics
to the middle in Fig.~\ref{fig:caus}, since the
amplification is sufficiently high between the peaks of the light curve.}
This result does not differ greatly from the result in Table
\ref{tab:MACHO9byMACHO}, 
which is a crosscheck of our fitting code.
The structure of the caustics and the trajectory of the source
are shown in Fig.~\ref{fig:static}.
The corresponding light curves are also shown.

\begin{figure}[t]
\begin{center}  
  \begin{tabular}[h]{ccc}
    \epsfysize 5cm 
    \epsfbox{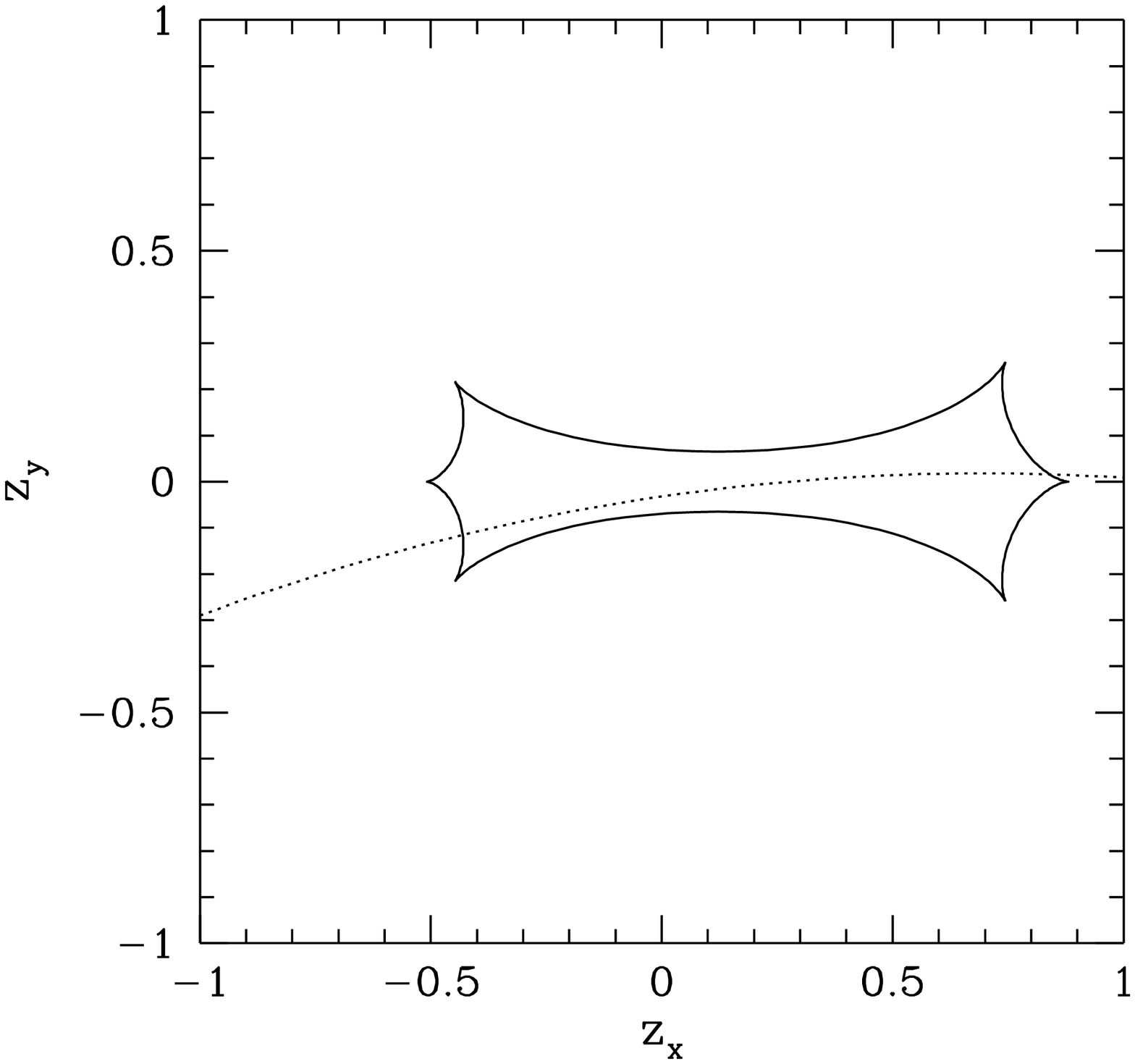} &
    \epsfysize 5cm 
    \epsfbox{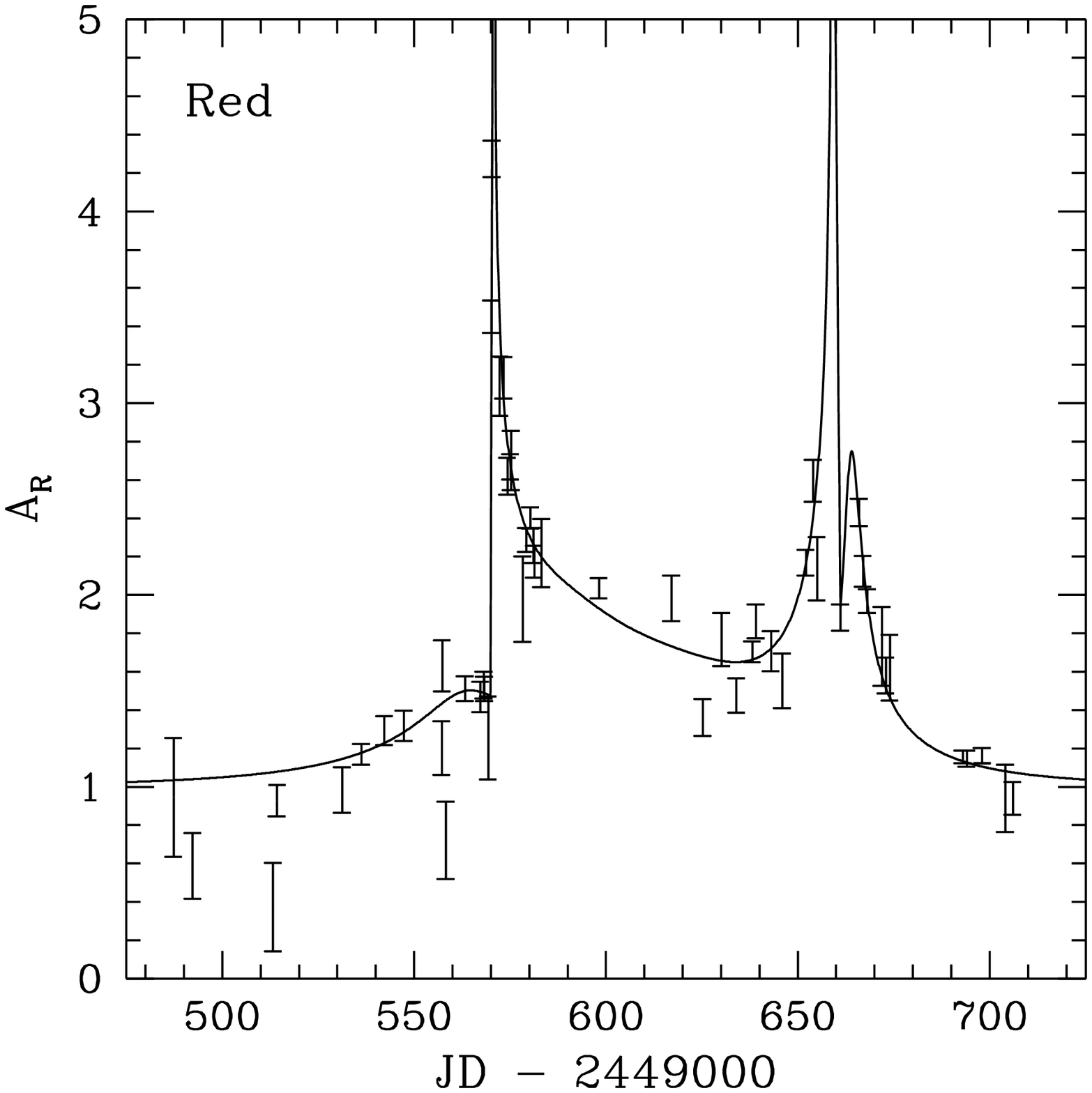} &
    \epsfysize 5cm 
    \epsfbox{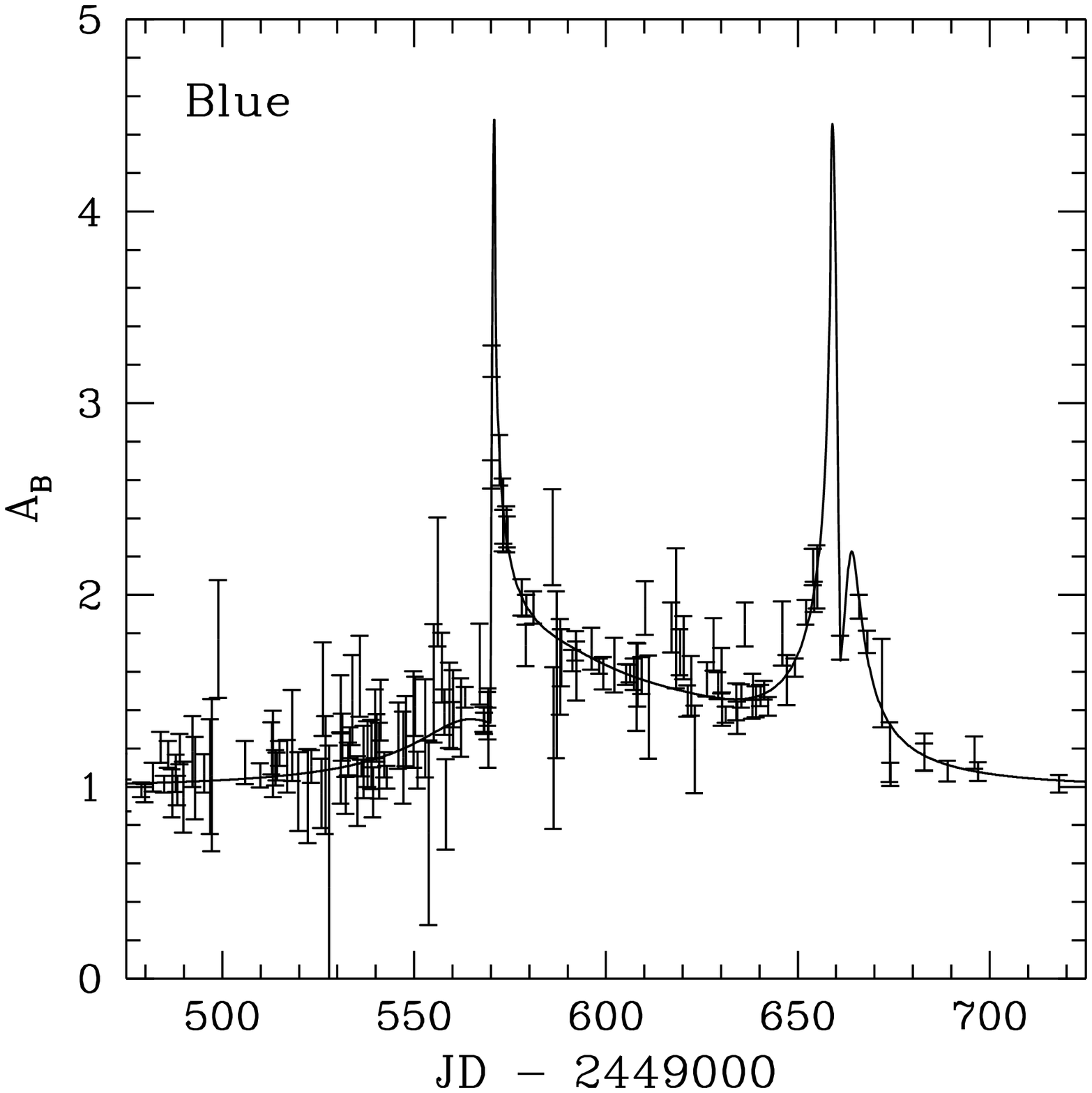}
  \end{tabular}
\end{center}    
  \caption{The structure of the caustics and the trajectory of the source
({\it left}).
The corresponding light curves for red ({\it middle})
and blue ({\it right}) are also shown.
The rotation effects are included.}
  \label{fig:rotate}
\end{figure}

Next we performed the fitting of the data of MACHO LMC-9
taking account of the
rotation of the lens binary.
The influence of the rotation of the lens binary on the light curve 
can be divided into two effects, the change of the caustics shape with time 
caused by the change of $l$ and rotation of the caustics with time 
in the lens plane, which is treated as a curve of the source trajectory. 
However, since the former effect is hardly separated from 
the ambiguity of the other parameters within the limited accuracy of the 
observation, we consider only the latter effect.
Thus, in order to evaluate the rotation effects,
it is useful to examine the case that the binary is face-on and $e=0$.
Therefore we assume that the binary is face-on and $e=0$ as a first step.
Then we need ten parameters, (1)-(10) in \S \ref{sec:para}.
The results of the $\chi^2$ fitting are shown in Table
\ref{tab:rotatingMACHO9}.
The structure of the caustics and the trajectory of the source are shown
in Fig.~\ref{fig:rotate}.
The corresponding light curves are also shown.

Contrary to our expectations, there are few differences between the
fitted parameters with and without the rotation 
from Tables \ref{tab:nonrotatingMACHO9} and \ref{tab:rotatingMACHO9}.
This is because
the period of the lens binary $t_B$ is quite large.
The binary rotates by only
$\sim 5.9^{\circ} (M/M_{\odot})^{1/4}$ 
during the Einstein radius crossing time.
Thus the rotation effects are very small.\footnote{
Of course, the fitted parameters may be at only a local minimum.
The period $t_B$ in the global minimum may be smaller.
However, it does not seem that $t_B$ is smaller,
since we usually find $t_B \simg 60$ even when we start the fitting
from $t_B \siml 60$.}
Note that the result of the small rotation effects
does not depend on the face-on and $e=0$ assumptions.
We can consider the following three possible reasons
for the contradiction of the simple estimate of $t_B$ in Eq.~(\ref{tBLMC1})
and the fitted value of $t_B$ in Table \ref{tab:rotatingMACHO9}:
\begin{enumerate}
\item The physical parameters in Eq.~(3.15), such as the mass $M$ and 
the velocity $V_{\bot}$, are not typical in this event.

\item The binary of this event is nearly edge-on.

\item The binary of this event is very eccentric.
\end{enumerate}
We consider the possible physical parameters
that account for the small rotation effects in \S \ref{sec:mvd},
we consider the possible inclination in \S \ref{sec:incli},
and we consider the possible eccentricity in \S \ref{sec:ecc}.

\subsection{The mass, the velocity and the distance of the lens}
\label{sec:mvd}
Assuming that the binary is face-on and that $e=0$,
we can obtain the mass, the velocity and the position of the lens
from only the fitted parameters
in Table \ref{tab:rotatingMACHO9} and the
radius of the source in Eq.~(\ref{defRstar}) as follows.\cite{dominik}
Note that the probability distribution of the transverse velocity
is not necessary.

\begin{figure}[t]
  \epsfysize 7cm \centerline{\epsfbox{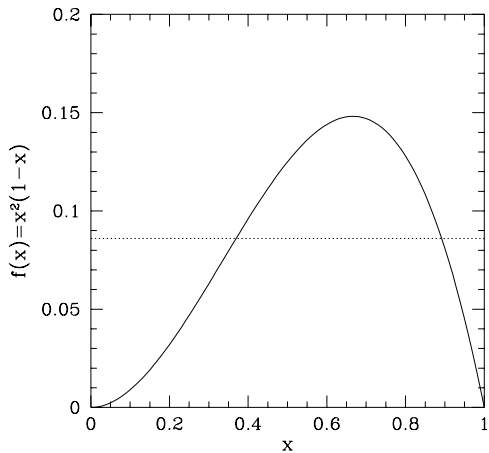}}
  \caption{$f(x)=x^2 (1-x)$.
The horizontal dotted line is the right-hand side of Eq.~($5\cdot3$) 
with the fitted parameters in Table III.}
  \label{fig:f(x)}
\end{figure}

We assume that the binary is face-on and that $e=0$.
The velocity of the lens $V_\bot$ is determined by
Eq.~(\ref{defTstar}) as
\begin{equation}
  V_\bot = x {{R_\star}\over{T_\star}}
  \label{defVbot}
\end{equation}
for given $x$.
Using Eqs.~(\ref{defEradi}), (\ref{deftE}) and (\ref{defVbot}),
the total mass of the lens $M$ is determined by
\begin{equation}
  M={{c^2 R_\star^2}\over{4 G D_{os}}}
  \left({{t_E}\over{T_\star}}\right)^2 {{x}\over{1-x}}
  \label{defM}
\end{equation}
for given $x$.
Using Eqs.~(\ref{defEradi}), (\ref{deftE}), (\ref{period}), (\ref{deftB})
and (\ref{defVbot}),
the position of the lens $x$ is obtained from
\begin{equation}
  x^2 (1-x) = {{t_B^2}\over{16\pi^2 a^3}}
  {{c^2 t_E T_\star}\over{D_{os} R_\star}}.
  \label{f(x)}
\end{equation}
The left-hand side of this equation, $f(x):=x^2 (1-x)$,
is plotted in Fig.~\ref{fig:f(x)}.
The function $f(x)$ has a maximum value $f(2/3)=4/27$
at $x=2/3$.
The equation $f(x)=y$ has two solutions, $x_1$ and $x_2$, for $0<y<4/27$.
The right-hand side of Eq.~(\ref{f(x)})
is determined from the fitted parameters.

After determining the positions, $x_1$ and $x_2$, with Eq.~(\ref{f(x)}),
we can obtain the mass and the velocity of the lens for each distance
with Eqs.~(\ref{defVbot}) and (\ref{defM}).
Since the right-hand side of Eq.~(\ref{f(x)}) is 0.0860
from Table \ref{tab:rotatingMACHO9},
two solutions of this equation are
\beq
x_1=0.369\quad{\rm and}\quad x_2=0.892.
\label{x1x2}
\eeq
The corresponding mass and velocity are
\beqa
M=0.000902\ M_{\odot},\quad V_\bot=7.30\ {\rm km/s}
\quad {\rm for}\ x=x_1,&&
\label{x1mass}\\
M=0.0127\ M_{\odot},\quad V_\bot=17.6\ {\rm km/s}\quad {\rm for}\ x=x_2.&&
\label{x2mass}
\eeqa
If the binary of the MACHO LMC-9 event has such physical parameters,
we can explain the small rotation effects.
However, these parameters seem to be quite strange.
The mass seems to be too small, and
the position favors the halo lens, while the transverse velocity
prefers the LMC lens.
However, we cannot draw definite conclusions from this event alone.

\subsection{Inclination}\label{sec:incli}
In this section we consider a possible inclination to explain
the small rotation effects in the MACHO LMC-9 event.
Without the face-on assumption, the fitted values of $t_B$ and $a$
do not generally coincide with the real values of $t_B$ and $a$.
We obtained the mass, the velocity
and the position using $\tilde t_B$ and $\tilde a$ instead of
$t_B$ and $a$, respectively, in Eq.~(\ref{f(x)}),
where $\tilde t_B$ and $\tilde a$ denote the fitted values of
$t_B$ and $a$, respectively.
Therefore a certain inclination may explain the small rotation
effects even if we use typical physical parameters.
To investigate the inclination effect, we set $e=0$.
Therefore the additional parameters are 
the Euler angles $\beta$ and $\gamma$, as shown in \S \ref{sec:para}.
Rigorously, the parameters $t_B$ and $a$ for given $\beta$ and $\gamma$
have to be determined by the fitting.
However, we can approximately determine $t_B$ and $a$ 
for given $\beta$ and $\gamma$
from $\tilde t_B$ and $\tilde a$ as follows.

Assuming that $\tilde a$ is mainly determined by the separation
between the projected masses at $t=t_S$,
the relation between $a$ and $\tilde a$ can be estimated by
Eq.~(\ref{proj1}) as
\beq
a={{\tilde a}\over{\sqrt{\cos^2\beta \cos^2\gamma + \sin^2 \gamma}}},
\label{reala}
\eeq
since $\phi=0$ at $t=t_S$ when $e=0$.
To determine the relation between $t_B$ and $\tilde t_B$,
we assume that the parameter $\tilde t_B$ is mainly determined
by the rotation angle projected on the lens plane.
In other wards, the relation between $t_B$ and $\tilde t_B$
is determined by the condition
that the projection of the rotation angle $4\pi/t_B$
between $t=t_S-1$ and $t=t_S+1$ coincides with $4\pi/\tilde t_B$.
The angle between the relative vector of the binary masses $\bm U$ and
the $x$-axis of the orbital plane at $t=t_S-1$ is $\phi_1:=-2\pi/t_B$.
The corresponding angle $\varphi_1$ between the relative vector projected
on the lens plane and the $x$-axis of the lens plane
is determined by Eq.~(\ref{proj2}) as
\beq
\tan \varphi_1 = {{\tan (\phi_1 + \gamma)}\over{\cos \beta}}.
\label{varphi1}
\eeq
Similarly, since the angle between the relative vector of the binary
masses $\bm U$ and the $x$-axis of the orbital plane at $t=t_S+1$ is
$\phi_2=2\pi/t_B$, the corresponding angle $\varphi_2$ between
the relative vector projected
on the lens plane and the $x$-axis of the lens plane
is determined by Eq.~(\ref{proj2}) as
\beq
\tan \varphi_2 = {{\tan (\phi_2 + \gamma)}\over{\cos \beta}}.
\label{varphi2}
\eeq
The relation between $t_B$ and $\tilde t_B$ is determined by the
condition $|\varphi_2 - \varphi_1| = 4\pi/\tilde t_B$.
With Eqs.~(\ref{varphi1}) and (\ref{varphi2}) this relation can be
obtained as
\beqa
{{2\pi}\over{t_B}}=\arctan
  \left[{{\mp\cos\beta +
        \sqrt{\cos^2\beta +
          \cos^4\gamma (\cos^2\beta + \tan^2\gamma)
          (\cos^2\beta \tan^2\gamma + 1)
          \tan^2(4\pi/\tilde t_B)}}
      \over{\cos^2\gamma (\cos^2\beta \tan^2\gamma + 1)
        \tan(4\pi/\tilde t_B)}}\right],
  \nonumber\\
\label{realtB1}
\eeqa
where the minus sign is for $0<\beta<\pi/2$ and the plus sign is for
$\pi/2<\beta<\pi$.

\begin{figure}[t]
  \epsfysize 7cm \centerline{\epsfbox{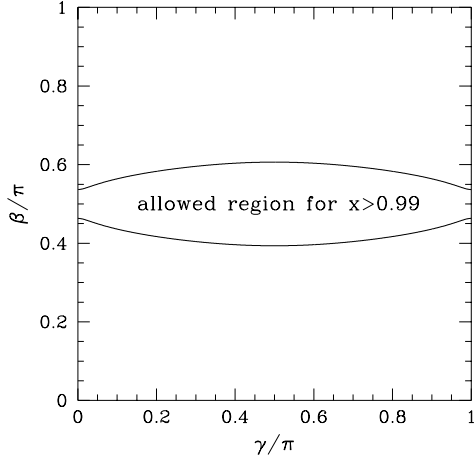}}
  \caption{The allowed region of the inclination $\beta$ and the phase 
$\gamma$ for $0.99<x<1$.}
  \label{fig:incli}
\end{figure}

Of course, we cannot determine $\beta$ and $\gamma$ by the fitted
parameters.
However, inversely, we can determine the allowed region of $\beta$ and
$\gamma$ for given $x$.
For example, if we assume that the lensing object resides in the LMC disk,
i.e.,
\beq
0.99<x<1,
\label{xregion}
\eeq
the left-hand side of Eq.~(\ref{f(x)}) is less than
$0.009801$.
Substituting Eqs.~(\ref{reala}) and (\ref{realtB1})
into Eq.~(\ref{f(x)}),
the allowed region of $\beta$ and $\gamma$ can be obtained
in Fig.~\ref{fig:incli}.\footnote{
We find that the inequalities $x<x_1$ and $x_2<x$
hold for any $\beta$ and $\gamma$,
since the inequality $t_B^2/a^3<\tilde t_B^2/\tilde a^3$ holds
irrespective of $\beta$ and $\gamma$.
However, these relations do not hold if we consider the eccentricity.}
The probability that the inclination and the phase
are in the allowed region is
$26.0\%$, assuming random inclination and phase.
The corresponding mass and velocity can be obtained 
by Eqs.~(\ref{defVbot}) and (\ref{defM}) as
\beqa
0.153\ {\rm} M_{\odot} < M,\quad
19.6\ {\rm km/s} < V_{\perp} < 19.8\ {\rm km/s}.
\label{mregion}
\eeqa

In this way, if the binary of this event is nearly edge-on,
we can explain the small rotation effects
with typical physical parameters.
However, we cannot draw definite conclusions from only this event.

\subsection{Eccentricity}\label{sec:ecc}
In this section we consider a possible eccentricity to explain
the small rotation effects in the MACHO LMC-9 event.
As in the previous section, without the $e=0$ assumption,
the fitted values $\tilde t_B$ and $\tilde a$
do not generally coincide with the real values of $t_B$ and $a$.
Since we use $\tilde t_B$ and $\tilde a$ instead of
$t_B$ and $a$ in Eq.~(\ref{f(x)}) 
to obtain the physical parameters,
a certain eccentricity may explain the small rotation
effects even if we use typical physical parameters.
%We consider the effect of the eccentricity in this section.
To investigate the eccentricity effect we consider the face-on binary.
Therefore, the additional parameters are 
$e$ and $T_0$, as shown in \S \ref{sec:para}.
We can approximately determine $t_B$ and $a$ for given $e$ and $T_0$
from $\tilde t_B$ and $\tilde a$ as follows.

Assuming that $\tilde a$ is mainly determined by the separation
between the projected masses at $t=t_S$,
the relation between $a$ and $\tilde a$ can be approximated by
\beq
a={{{\tilde a} (1+e \cos \phi_3)}\over{1-e^2}},
\label{reala2}
\eeq
from Eqs.~(\ref{blength}) and (\ref{bangle}),
where $\phi_3$ is determined from Eqs.~(\ref{btime}) and (\ref{bangle})
with $t=t_S$.
To determine the relation between $t_B$ and $\tilde t_B$,
we assume that the parameter $\tilde t_B$ is mainly determined
by the rotation angle projected on the lens plane.
In other words, the relation between $t_B$ and $\tilde t_B$
is determined by the condition
that the rotation angle between $t=t_S-1$ and $t=t_S+1$ 
coincide with $4\pi/\tilde t_B$.
The angle $\phi_4$ between the relative vector between the masses
and the $x$-axis of the lens plane at $t=t_S-1$ is
determined by
\beqa
t_S-1-t_0&=&{{t_B}\over{2\pi}}(\lambda_4-e\sin\lambda_4),
\label{chi4}\\
\cos\phi_4&=&{{\cos\lambda_4-e}\over{1-e\cos\lambda_4}},
\label{phi4}
\eeqa
from Eqs.~(\ref{btime}) and (\ref{bangle}).
Similarly, the angle $\phi_5$ between the relative vector between the masses
and the $x$-axis of the lens plane at $t=t_S+1$ is
determined by
\beqa
t_S+1-t_0&=&{{t_B}\over{2\pi}}(\lambda_5-e\sin\lambda_5),
\label{chi5}\\
\cos\phi_5&=&{{\cos\lambda_5-e}\over{1-e\cos\lambda_5}}.
\label{phi5}
\eeqa
The relation between $t_B$ and $\tilde t_B$ is determined by the
condition $\phi_5 - \phi_4 = 4\pi/\tilde t_B$.
Subtracting Eq.~(\ref{chi4}) from Eq.~(\ref{chi5}),
we can determine $t_B$ as
\beq
{{2\pi}\over{t_B}} = Y - e \cos X \sin Y,
\label{realtB2}
\eeq
where $X:=(\lambda_4+\lambda_5)/2$ and $Y:=(\lambda_5-\lambda_4)/2$
are determined by
\beq
t_S-t_0={{X - e \sin X \cos Y}\over{Y - e \cos X \sin Y}},
\eeq
from Eqs.~(\ref{chi4}) and (\ref{chi5}), and
\beq
\cos \left({{4\pi}\over{\tilde t_B}}\right)
= {{e^2 \cos^2 X + (2 - e^2) \cos^2 Y - 2 e \cos X \cos Y + e^2 - 1}\over
{e^2 (\cos^2 X + \cos^2 Y) - 2 e \cos X \cos Y + 1 - e^2}},
\eeq
from Eqs.~(\ref{phi4}) and (\ref{phi5}) and the condition 
$\phi_5 - \phi_4 = 4\pi/\tilde t_B$.

\begin{figure}[t]
  \epsfysize 7cm \centerline{\epsfbox{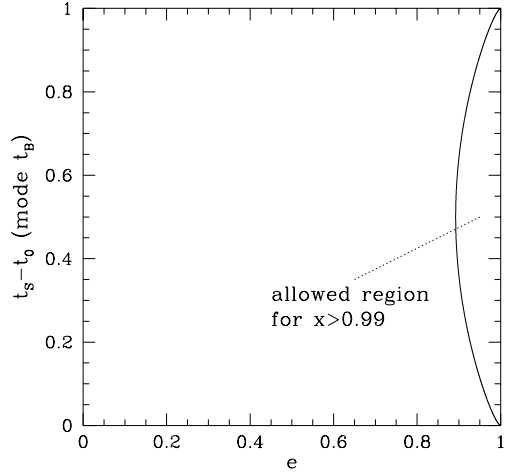}}
  \caption{The allowed region of the eccentricity $e$ and 
$t_S-t_0\ ({\rm mode}\ t_B)$ for $0.99<x<1$.}
  \label{fig:ecc}
\end{figure}

We assume that the lensing object resides in the LMC 
in Eq.~(\ref{xregion}).
Substituting Eqs.~(\ref{reala2}) and (\ref{realtB2})
into Eq.(\ref{f(x)}),
the allowed region of $e$ and $t_S-t_0\ ({\rm mode}\ t_B)$ can be obtained
in Fig.~\ref{fig:ecc}.
The minimum eccentricity of this allowed region is 0.892.
The corresponding mass and velocity are in Eq.~(\ref{mregion}).
In this way, if the binary of this event is very eccentric,
we can explain the small rotation effects
with typical physical parameters.
However we cannot say for certain from only this event.

\section{Summary and discussion}\label{sec:summary}
We investigated the rotation effects of lens binaries on 
the binary-microlensing events towards the LMC and SMC.
It is found that the rotation effects cannot always be
neglected when the lens binaries are in the LMC disk or the SMC disk,
while the rotation can be neglected when the lenses are in our halo.
%in \S \ref{sec:rote}.
It follows from this that we need to consider the rotation effects
in the analyses of the coming binary events
if the microlensing events towards the Magellanic Clouds are self-lensing.

In the MACHO LMC-9 event the transverse velocity
prefers the lens binary in the LMC disk.\cite{bennett}
For this reason, we reexamined this event in detail as an example.
The simple estimate
in Eq.~(\ref{tB9}) indicates that the binary
rotates by more than $\sim 60^{\circ} (M/M_{\odot})^{1/4}$ 
during the Einstein radius crossing time.
Therefore we cannot neglect the rotation of the lens binary
in the fitting of MACHO LMC-9.
We performed the fitting of MACHO LMC-9 taking the rotation into account
on the assumption that the binary is face-on and that $e=0$.

Contrary to our expectation, 
the rotation effects are small,
i.e., the projected rotation angle is
only $\sim 5.9^{\circ} (M/M_{\odot})^{1/4}$ 
during the Einstein radius crossing time.
Note that this result does not depend on the face-on and $e=0$ assumptions.
We can consider three possible reasons
for this contradiction between the simple estimate of $t_B$ in Eq.~(\ref{tB9})
and the fitted value of $t_B$ in Table \ref{tab:rotatingMACHO9}.
The first possibility is that 
the physical parameters in Eq.~(\ref{tB9}) are different in this event
as in Eqs.~(\ref{x1x2})-(\ref{x2mass}),
although these parameters are quite peculiar.
The second possibility is that the binary of this event is nearly edge-on
with typical physical parameters $0.99<x<1$ and Eq.(\ref{mregion}), 
as shown in Fig.~\ref{fig:incli},
although the probability of $\beta$ and $\gamma$ being in the allowed region 
is only $26\%$ assuming random inclination and phase.
The third possibility is that the binary of this event is very eccentric
with typical physical parameters $0.99<x<1$ and Eq.~(\ref{mregion}), 
as shown in Fig.~\ref{fig:ecc},
although the minimum eccentricity of the allowed region is 0.892.
However, since we cannot determine all physical parameters
including the inclination and the eccentricity
only from the light curve,
we cannnot draw definite conclusions from only this event.

The transverse velocity is somewhat
smaller than that expected for a lens in the LMC disk.
This result is the same as that of the analysis 
of Bennett et al.,\cite{bennett}
even if we take the rotation of the lens binary into account.
Since the incident angle of the trajectory of the source into the
caustics is not greatly changed,
the transverse velocity remains small.

The fitted parameters only represent a local minimum.
However, the complete analysis is difficult, since
the number of data points around the peak of the light curve
is so small that there will be many local minima.\cite{ALB99}
The lack of data also allows the possibility that the source is a binary,
as Bennett et al. \cite{bennett} claimed.
We agree that it is important to 
determine whether the source is a binary or not with HST.
In order not to overlook the important information 
from the caustic crossing events,
it is necessary to monitor the events frequently,
such as with the MOA collaboration \cite{MU99} and the Alert system.

Note that the rotation effects are not always important
when the lens is in the LMC disk or the SMC disk.
For example, in the MACHO 98-SMC-1 event the transverse velocity
is larger than $\sim 60$ km/s.\cite{AF98,ALB98,ALC98}
Since $t_B \simg 20 \gg 1$ from Eq.~(\ref{tBLMC1}),
the rotation effects are not efficient in this event.
Our statement is that there is a {\it sufficient probability}
that the rotation effects are important for self-lensing.

If the microlensing events towards the halo
are due to self-lensing,\cite{sahu,wu}
the number of binary-events for which the rotation effects are important 
will increase.
We will be able to crosscheck
whether or not the microlensing events are self-lensing
by examining the rotation effects,
together with the transverse velocity distribution.
As the number of binary-lens events increases,
stronger constraints on the nature of the lenses will be obtained
statistically.

\section*{Acknowledgements}
We would like to thank D. Bennett and A. Becker for providing 
the data of the MACHO LMC-9 event.
We would also like to thank H. Sato, T. Nakamura
and M. Siino for continuous encouragement and useful discussions.
We are also grateful to N. Sugiyama for carefully reading and commenting on
the manuscript.
We are also grateful to an anonymous referee 
for useful comments that helped improve the paper.
Numerical computation in this work was carried out in part
at the Yukawa Institute Computer Facility.
This work was supported in part by
Grant-in-Aid for Scientific Research of the Ministry of Education,
Science, Sports, and Culture of Japan, No. 08740170 (RN) and No. 9627 (KI),
and by the Japanese Grant-in-Aid 
for Scientific Research on Priority Areas, No. 10147105 (RN).

\appendix
\section{Notations} %Empty argument \section{} yields `Appendix'.
\begin{namelist}{${\bm U}_1,\ {\bm U}_2$   }
\item[${\bfxi}$] : image position in the lens plane
\item[${\bfeta}$] : source position in the source plane
\item[${\bfxi}_1,\ {\bfxi}_2$] : positions of the masses of the lens binary
  in the lens plane
\item[${\bm r}$] : image position in the lens plane in units of $\rho_E$
\item[${\bm z}$] : source position in the source plane in units of $\rho_E$
\item[${\bm r}_1,\ {\bm r}_2$] : positions of the masses of the lens
  binary in the lens plane in units of $\rho_E$
\item[${\bm U}$] : relative vector between the lens masses in the
  orbital plane
\item[${\bm U}_1,\ {\bm U}_2$] : position of the masses of the lens
  binary in the orbital plane
\item[${\bm u}$] : relative vector between the lens masses
  in units of $\rho_E$
\item[$M_1,\ M_2$] : masses of the lens binary
\item[$M$] : $M_1+M_2$, total mass of the lens binary
\item[$\mu_i$] : $M_i/M$, mass fraction ($i=1,2$)
\item[$I$] : amplification factor  
\item[$f_{oR}$] : fraction of the lensed brightness of the lensed star
  in the red band
\item[$f_{oB}$] : fraction of the lensed brightness of the lensed star
  in the blue band
\item[$D_{ol}$] : distance between the observer and the lens plane
\item[$D_{os}$] : distance between the observer and the source plane
\item[$D_{ls}$] : distance between the lens plane and the source plane
\item[$x$] : ${D_{ol}}/{D_{os}}$
\item[$\rho_E$] : Einstein radius
\item[$V_{\bot}$] : transverse velocity of the source in the lens
  plane
\item[$V_{T}$] : transverse velocity of the source in the source
  plane
\item[$b$] : impact parameter of the source in
  the lens plane in units of $\rho_E$
\item[$\theta$] : angle between the $x$-axis
  and the trajectory of the source
\item[$A$] : semimajor axis of the lens binary
\item[$a$] : semimajor axis of the lens binary in units of $\rho_E$
\item[$\tilde a$] : fitted value of $a$ assuming $\beta=0$ and $e=0$
\item[$l$] : separation of the lens masses projected on the lens plane in
units of $\rho_E$
\item[$e$] : eccentricity of the lens binary
\item[$\phi$] : azimuthal angle of the binary in the orbital plane
\item[$\varphi$] : angle between the relative vector projected 
on the lens plane and the $x$-axis of the lens plane
\item[$t_E$] : Einstein radius crossing time
\item[$t$] : $T/t_E$, time in units of $t_E$
\item[$T_S$] : time when the source most closely approaches the origin
\item[$t_S$] : $T_S/t_E$, time when the source most closely approaches
  the origin in units of $t_E$
\item[$T_B$] : period of the lens binary
\item[$t_B$] : $T_B/t_E$, period of the lens binary in units of $t_E$
\item[$\tilde t_B$] : fitted value of $t_B$ assuming $\beta=0$ and $e=0$
\item[$T_0$] : time when the binary is at the pericenter
\item[$t_0$] : $T_0/t_E$, time when the binary is at the pericenter in units
  of $t_E$
\item[$T_\star$] : time for the source to move by one stellar radius of
  the source
\item[$R_\star$] : radius of the source
\item[$\alpha,\beta,\gamma$] : Euler angles which determine the orbital plane
(see Fig.~\ref{fig:euler})
\end{namelist}

\section{Rotation Effects of a Close Binary}\label{sec:close}
In this appendix, we show that the rotation effects of a typical MACHO
binary are small even when the binary is the close one.
For a close binary (i.e. $a \ll 1$), it seems that the effects of the
rotation can be large from Eq.~(\ref{halotB}).
For example, $t_B$ is about $4.2$ when $a$ is 0.1,\footnote{
Note that the probability of the event being observed as the
microlensing by two point masses decreases when the semimajor axis
becomes small.\cite{gaudigould}}
which implies that the binary rotates by about $85^{\circ}$ during the
Einstein radius crossing time.
However, we have to note that the region where the rotation effects
are important is not within the Einstein radius
but within the caustics.
This is because far from the caustics, the light curve due to two point masses
is almost the same
as that due to a point mass lens.
Therefore it is not reasonable to compare the binary period with the
Einstein radius crossing time in the estimate of the rotation effects
in Eq.~(\ref{halotB}).
Instead, we have to compare the binary period
with the caustics crossing time.

To treat the problem analytically, we assume that the mass ratio is
unity: $\mu_1=\mu_2=1/2$.
When the semimajor axis is small (i.e. $a \to 0$),
the size of the caustics $z_{\rm cau}$ satisfies \cite{schneider}
\begin{equation}
  z_{\rm cau} \sim {{a^2}\over{2}}.
\end{equation}
Thus, when the semimajor axis $a$ is small, the ratio of the binary period to
the caustics crossing time, $t_B^{\rm cau}$, is approximately
\beqa
t_B^{\rm cau} &\sim& {{T_B}\over{z_{\rm cau} t_E}}
\sim {{2}\over{a^2}} t_B
\nonumber\\
&\sim& 2\times 134 \left({{a}\over{1}}\right)^{-1/2}
\left( {{D_{OS}}\over{50\ {\rm kpc}}} \right)^{1/4}
\left({{M}\over{M_{\odot}}}\right)^{-1/4}
\left({{V_{\perp}}\over{200\ {\rm km/s}}}\right)
\left({{x(1-x)}\over{0.5(1-0.5)}}\right)^{1/4}
\nonumber\\
&\propto& a^{-1/2}
\eeqa
for a typical MACHO.
The exponent with respect to the semimajor axis $a$ 
changes from $3/2$ to $-1/2$.
Therefore the rotation effects are small even when the binary is
the close one, as long as the binary is in the Milky Way halo.

\begin{table}[b]
  \caption{Fitted parameters for MACHO LMC-9 
       	from Bennett et al. (1996).
The rotation effects are not included.}
  \label{tab:MACHO9byMACHO}
  \begin{center}
    \begin{tabular}{l|r}
%      \noalign{\hrule height0.8pt}
	\hline
	\hline
      parameters & fitted values \\
      \hline
      $f_{oR}$ & $0.259 \pm 0.002$\\
      $f_{oB}$ & $0.174 \pm 0.001$\\
      $t_{E}$ [days]& $71.7 \pm 0.1$ \\
      $T_S$ [days]& $603.04 \pm 0.02$\\
      $b$ & $-0.055 \pm 0.001$\\
      $\theta$ [radian]& $0.086 \pm 0.001$\\
      $\mu_1$ & $0.620 \pm 0.002$\\
      $l$ & $1.6545 \pm 0.0008$\\
      $T_\star$ [days]& $0.65 \pm 0.18$ \\
      \hline
      $\chi^2$(for 848 degrees) & $1489$ \\
      reduced $\chi^2$ & $1.76$\\
      \noalign{\hrule height0.8pt}
    \end{tabular}
  \end{center}
\end{table}

\begin{table}[b]
  \caption{Fitted parameters for MACHO LMC-9 from our 
      fitting code.
The rotation effects are not included.}
  \label{tab:nonrotatingMACHO9}
  \begin{center}
    \begin{tabular}{l|r}
%      \noalign{\hrule height0.8pt}
	\hline
	\hline
      parameters & fitted values \\
      \hline
      $f_{oR}$ & $0.255$ \\
      $f_{oB}$ & $0.176$ \\
      $t_{E}$ [days]& $71.4$ \\
      $T_S$ [days]& $603.1$ \\
      $b$ & $-0.0538$ \\
      $\theta$ [radian]& $0.0843$ \\
      $\mu_1$ & $0.617$ \\
      $a$ & $1.66$ \\
      $T_\star$ [days]& $0.611$ \\
      \hline
      $\chi^2$(for $871-9$ degrees) & $1396$ \\
      reduced $\chi^2$ & $1.62$\\
      \noalign{\hrule height0.8pt}
    \end{tabular}
  \end{center}
\end{table}

\begin{table}[b]
  \caption{Fitted parameters for MACHO LMC-9 from our fitting code.
The rotation effects are included.}
  \label{tab:rotatingMACHO9}
  \begin{center}
    \begin{tabular}{l|r}
%      \noalign{\hrule height0.8pt}
	\hline
	\hline
      parameters & fitted values \\
      \hline
      $f_{oR}$ & $0.246$ \\
      $f_{oB}$ & $0.172$ \\
      $t_{E}$ [days]& $69.4$ \\
      $T_S$ [days]& $601.4$ \\
      $b$ & $-0.0319$ \\
      $\theta$ [radian]& $0.144$ \\
      $\mu_1$ & $0.560$ \\
      $a$ & $1.69$ \\
      $T_\star$ [days]& $0.611$ \\
      $t_B$ & $60.6$\\
      \hline
      $\chi^2$(for $871-10$ degrees) & $1372$ \\
      reduced $\chi^2$ & $1.59$\\
      \noalign{\hrule height0.8pt}
    \end{tabular}
  \end{center}
\end{table}

\end{document}